\documentclass[review]{elsarticle}

\usepackage{lineno,hyperref}
\modulolinenumbers[5]

\journal{Elsevier Journal}

\usepackage[utf8]{inputenc} 
\usepackage[T1]{fontenc}    
\usepackage{hyperref}       
\usepackage{url}            
\usepackage{booktabs}       
\usepackage{amsfonts}       
\usepackage{nicefrac}       
\usepackage{microtype}      
\usepackage{lipsum}
\usepackage{color,soul}     
\usepackage{amsmath}        
\usepackage{graphicx}       

\usepackage{subfig} 

\usepackage{multirow}

\usepackage{setspace}
\usepackage{subfig} 

\usepackage{nomencl}
\renewcommand{\nomname}
\makenomenclature

\usepackage[version=4]{mhchem}

\DeclareMathOperator{\arcsinh}{arcsinh}
\renewcommand{\vec}[1]{\boldsymbol{#1}}

\usepackage{color,soul}

\begin{document}

\begin{frontmatter}

\title{Enhanced physics-constrained deep neural networks for modeling vanadium redox flow battery}


\author[PNNL,UMN]{QiZhi He}
\ead{qzhe@umn.edu}


\author[PNNL]{Yucheng Fu}
\ead{yucheng.fu@pnnl.gov}

\author[PNNL]{Panos Stinis}
\ead{panagiotis.stinis@pnnl.gov}

\author[UIUC,PNNL]{Alexandre Tartakovsky} 
\ead{amt1998@illinois.edu}

\address[PNNL]{Physical and Computational Sciences Directorate, Pacific Northwest National Laboratory Richland, WA 99354}
\address[UIUC]{Department of Civil and Environmental Engineering, University of Illinois Urbana-Champaign, Urbana, IL 61801}

\address[UMN]{Department of Civil, Environmental, and Geo-Engineering, University of Minnesota, Minneapolis, MN 55455}

\begin{abstract}
Numerical modeling and simulation have become indispensable tools for advancing a comprehensive understanding of the underlying mechanisms and cost-effective process optimization and control of flow batteries.
In this study, we propose an enhanced version of the physics-constrained deep neural network (PCDNN) approach \cite{he2021physics} to provide high-accuracy voltage predictions in the vanadium redox flow batteries (VRFBs).
The purpose of the PCDNN approach is to enforce the physics-based zero-dimensional (0D) VRFB model in a neural network to assure model generalization for various battery operation conditions. Limited by the simplifications of the 0D model, the PCDNN cannot capture sharp voltage changes in the extreme SOC regions. To improve the accuracy of voltage prediction at extreme ranges, we introduce a second (enhanced) DNN to mitigate the prediction errors carried from the 0D model itself and call the resulting approach enhanced PCDNN (ePCDNN). By comparing the model prediction with experimental data, we demonstrate that the ePCDNN approach can accurately capture the voltage response throughout the charge--discharge cycle, including the tail region of the voltage discharge curve. Compared to the standard PCDNN, the prediction accuracy of the ePCDNN is significantly improved. The loss function for training the ePCDNN is designed to be flexible by adjusting the weights of the physics-constrained DNN and the enhanced DNN. This allows the ePCDNN framework to be transferable to battery systems with variable physical model fidelity.
\end{abstract}


\end{frontmatter}


\section{Introduction}
%

To achieve carbon neutrality, utilization of renewable energy sources has increased to mitigate the $\text{CO}_{2}$ emissions from fossil fuels \cite{chen2021carbon}.  To store the harvested energy and provide uninterrupted power grid supply, the large-scale battery storage system can serve as a critical component that supports renewable energy development. Among existing technologies, the redox flow battery (RFB) is one of the most promising candidates due to its high energy capacity and rapid response to grid supply demand \cite{soloveichik2015flow,noack2015chemistry,weber2011redox}. In RFBs, the positive and negative electrolytes are stored in external tanks separate from the electrodes. During operation, pumps are used to circulate the electrolytes to the battery cell electrodes for reactions to occur. These features make RFBs relatively safe, and the adjustable storage tank size can easily accommodate variable demand for battery capacity.  By separating the anolyte and catholyte, the problem of self-discharge during prolonged storage periods can also be mitigated \cite{tokuda1998development}.  

The existing RFBs contain a large selection of redox couples, including $\text{Cr}^{2+}\slash \text{Cr}^{3+} vs. \text{Fe}^{2+}\slash \text{Fe}^{3+}$\cite{lopez1992optimization}, $\text{V}^{2+}/\text{V}^{3+} vs. \text{Br}^{-}/\text{ClBr}_2$ \cite{skyllas2003novel,skyllas2004kinetics}, $\text{Fe}^{2+}/\text{Fe}^{3+} vs. \text{Ti}^{2+}/\text{Ti}^{4+}$ \cite{wang1984study}, etc.  The all-vanadium redox flow battery (VRFB) utilizes a single vanadium element for both the negative and positive cells. The $\text{V}^{2+}/\text{V}^{3+}$ redox couple forms the anolyte, and the $\text{V}^{4+}(\text{VO}^{2+})/\text{V}^{5+}(\text{VO}^+_2)$ redox couple forms the catholyte.  With full vanadium species, the cross-transport of active species can be well controlled during operation \cite{wang2013recent}. In past decades, extensive experimental and modeling investigations have been carried out to acquire a more comprehensive understanding of VFRBs.  The cell/electrode design, cycle life, and power efficiencies have been continuously optimized \cite{rychcik1988characteristics,luo2008preparation, jiang2020high}, and capital cost has been driven down \cite{skyllas2019performance}.  In the current work, we select a VRFB system to demonstrate the development of an enhanced physics-constrained deep neural network (ePCDNN) for accurately predicting battery performance, including the extremes of the state of charge (SOC) vs. Voltage curve.

To enforce physics on a DNN, a VRFB model should be selected to constrain the ePCDNN. The existing VRFB models can be divided into zero-dimensional (0D), one-dimensional (1D), two-dimensional(2D), and three-dimensional(3D) categories based on the number of spatial dimensions considered for the electrochemical species concentration. The analytical models mostly fall into the 0D, 1D, and 2D categories\cite{Chen2021}. For numerical models using finite elements of finite difference methods, the complex VRFB cell design can be explored by considering electrochemical and species variation in the 3D space.

The 0D model simplifies the VRFB battery by representing the spatially related characteristics through a spatially averaged parameter. Recent studies have continuously improved the 0D model and yield satisfactory voltage predictions \cite{You2009a,Shah2011,Chen2014a,Eapen2019}. This provides a fast and cost-effective way to predict VRFB performance, making it suitable for real-time monitoring and control-related applications. One of the challenges for the 0D model is that the physical- and chemical-related model parameters must be very carefully measured or calibrated with experimental data to achieve satisfactory prediction accuracy. Given the fast pace of new electrode~\cite{zhou2019nano,mayrhuber2014laser} and electrolyte~\cite{li2021symmetry} development, battery performance prediction using the 0D model becomes more challenging because these parameters are unknown for new VRFB cells.


In a recent study, He et al. ~\cite{he2021physics} proposed a novel approach for estimating the parameters of the 0D model that represents them as functions of the operating conditions using a DNN. The estimated parameters are used in the 0D model to predict the voltage, and discrepancies between predictions and measurements are used to formulate a loss function for training the DNN. We refer to this approach as a physics-constrained deep neural network (PCDNN). In the PCDNN, four typical unknown parameters associated with the 0D model, including $\sigma_e$, $S$, $k_n$, and $k_p$, are estimated for each different operation condition. It has been demonstrated that the PCDNN approach can achieve more accurate parameter estimation than the least squares estimation scheme (LSE) and also provides better predictive generalization than approaches using the baseline parameters from literature \cite{he2021physics}. However, limited by the simplifications and assumptions in a 0D model, voltage prediction at extreme ranges (i.e., the VRFB state associated with very small or large SOCs) still cannot be well captured. Although the tail region discrepancy contributes a small portion to the overall RMSE of voltage prediction, the accurate prediction of the voltage curve at extremes is important for VRFB characterization and design optimization. For example, the cut-off time of the discharge can be accurately determined only through accurate voltage prediction at the end of the discharge when the SOC is small. This, in turn, can protect the VRFB from side effects (mixing, gas releasing, etc.).  In other systems, such as Ni-MH batteries, proper selection of cut-off times can affect the battery life cycle \cite{LIU2005270}.

\begin{figure}[htb!]
	\centering
	\includegraphics[angle=0,width=3in]{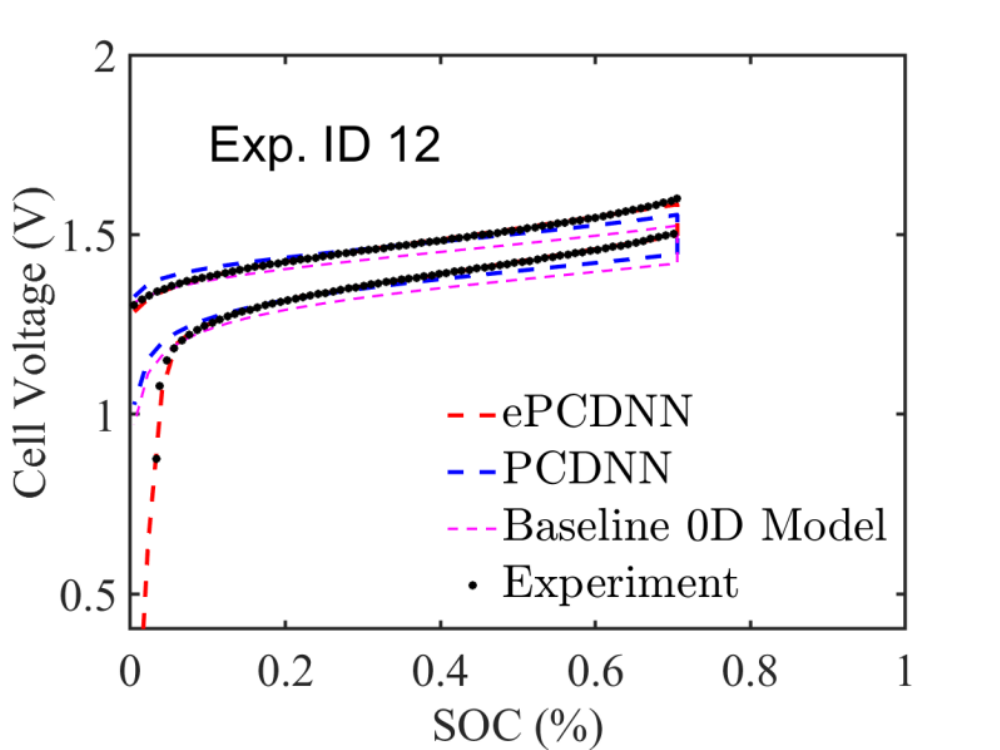}
	\caption{The comparison of SOC--voltage predictions for an exemplary experiment (Case 12 in Table \ref{table:PNNL_exp}) using the proposed approach (ePCDNN) and PCDNN. The VRFB 0D model (the purple dashed line) using the baseline parameters in Table \ref{table:exp_model_para} is also provided.}
	\label{fig:exp_comp_ID12}
\end{figure}

In the current work, we develop an ePCDNN model for a VRFB by leveraging both experimental observations and a physics-based computational model. Compared to PCDNN, the ePCDNN exhibits significantly improved cell voltage prediction accuracy at the extreme regions of the charge--discharge cycle.  This can improve precise determination of battery cut-off times for charge or discharge to prevent severe side effects during battery operation. Fig. \ref{fig:exp_comp_ID12} summarizes the comparison of the proposed ePCDNN, PCDNN, and 0D models. This experiment is the $12$th case shown in Table \ref{table:PNNL_exp}, and the baseline result is provided by using the 0D model with the calibrated parameters \cite{he2021physics,cheng2020data} given in Table \ref{table:exp_model_para}. The black dots stand for the experimental voltage data of the measured VRFB. During the charging period, the cut-off voltage is generally set to 1.5--1.6 V with a voltage span of 0.2--0.3V in the SOC range of [0, 0.7]. For the discharging period, the cut-off voltage is generally set to 0.6--0.8 V for a continuous discharging process with moderate current density.  The dominant voltage drop takes place near the end of the discharge period, with a small SOC. The sharp tail creates extra difficulties for voltage prediction. As can be seen in the figure, the PCDNN (the blue dashed line) has relatively good voltage prediction accuracy at large SOC compared to the baseline 0D model (the pink dash line) using parameter values from the literature.  At the end of discharge, where the SOC value is small (the "tail region"), the PCDNN and baseline models are discrepant from the experimental measurements. With the ePCDNN, the sharp voltage drop in the tail region can be accurately captured with an enhanced DNN to compensate for the limitations of the PCDNN model.  

The paper is organized as follows. A brief presentation of the implemented 0D VRFB model and experimental data can be found in Section~\ref{sec:model_0D} and Section~\ref{sec:exp_data}, respectively. The mathematical description of the ePCDNN is provided in Section~\ref{sec:Method}. Section~\ref{sec:Results} contains numerical results. We provide conclusions and a discussion of future work in Section~\ref{sec:Conclusions}.  



\section{Numerical VRFB model}\label{sec:model_0D}

%

A zero-dimensional (0D) numerical model has been developed in \cite{Shah2011,Sharma2015,Eapen2019,Lee2019c} to describe the governing physics and electrochemical reactions in the VRFB by including the components of the electrolytes, electrodes, and membrane. The 0D model lumps the variables in each VRFB component by assuming that the related quantities, such as concentration and voltage potentials, are uniform under the isothermal operation conditions.  By eliminating spatial dependency, the 0D model can provide fast and real-time predictions, which make it suitable for integration into neural networks. This section reviews the critical equations in the 0D model to enforce physics in the neural networks, and more details can be found in previous work\cite{he2021physics}. In this work, the active species studied in the electrolyte include $\mathcal{S} = \{\text{V(II)}, \text{V(III)}, \text{V(IV)}, \text{V(V)}, \text{H}^+, \text{H}_2 \text{O}\}$, and $c_i (i \in \mathcal{S})$ denotes the concentration of the species $i$. 
With a uniform flow rate $\omega$, the time-dependent analytical solution of species concentrations can be derived and summarized in Table \ref{table:conc_soln}. 

\begin{table}[htb]
	\centering
	\small{
		\caption{Analytical solutions of the concentrations of species in the VRFB cell \cite{he2021physics}.}
		\label{table:conc_soln}
		\scalebox{1.0}{
			\begin{tabular}{cc}
				\toprule
				Species $i$		&  Equation of the concentration $c_i$ \\
				\hline
				$ \text{V(II) or V(V)} $ & $c_i = c_i^0 - \frac{I}{V_e F \epsilon \tilde{ \epsilon} } \left(\frac{\epsilon \delta + e^{-\tilde{ \epsilon} t}}{1+ \epsilon \delta} - 1 - \frac{\epsilon \delta}{\tau} t \right)$ \\
				$ \text{V(III) or V(IV)} $ & $ c_i = c_i^0 + \frac{I}{V_e F \epsilon \tilde{ \epsilon}} \left(\frac{\epsilon \delta + e^{-\tilde{ \epsilon} t}}{1+ \epsilon \delta} - 1 - \frac{\epsilon \delta}{\tau} t \right)$ \\
				\hline
				$ \text{H}_2 \text{O}$ in negative electrode & $c_{i} = c_{i}^0 
				- \frac{I n_d}{V_e F \epsilon \tilde{ \epsilon} }
				\left(\frac{\epsilon \delta + e^{-\tilde{ \epsilon} t}}{1+ \epsilon \delta} - 1 - \frac{\epsilon \delta}{\tau} t \right)$ \\
				$ \text{H}_2 \text{O}$ in positive electrode & $c_{i} = c_{i}^0 
				+ \frac{I}{V_e F \epsilon \tilde{ \epsilon}} (1+n_d)
				\left(\frac{\epsilon \delta + e^{-\tilde{ \epsilon} t}}{1+ \epsilon \delta} - 1 - \frac{\epsilon \delta}{\tau} t \right)$ \\		
				\hline		
				$ \text{H}^+$ in positive electrode & $c_{i} = c_{i}^0
				- \frac{I}{V_e F \epsilon \tilde{ \epsilon} }
				\left(\frac{\epsilon \delta + e^{-\tilde{ \epsilon} t}}{1+ \epsilon \delta} - 1 - \frac{\epsilon \delta}{\tau} t \right)$ \\
				\bottomrule
			\end{tabular}%
		}
	}
\end{table}

In the equations listed in Table \ref{table:conc_soln}, $c_i^0$ is the initial concentration of the respective species, $I$ is the applied current, $F$ is the Faraday coefficient, $n_d$ is the drag coefficient, and $\epsilon$ and $V_e$ are the porosity and volume of the electrode, respectively.
The breadth, width, and length of the electrode are denoted as $b_e$, $w_e$, and $h_e$, respectively.
The electrolyte flow velocity (in $\text{m/s}$) is defined as $u = \omega/A_{in} \epsilon$, and the average electrolyte flow velocity $\tilde{u}$ in the porous medium is calculated as $\epsilon u$, where $A_{in} = b_e w_e$ is the inlet area of the electrode.
For the other symbols, $\delta = V_e/V_r$ is the ratio of the volumes of the electrode and reservoir, $\tau = h_e/u$, and $\tilde{\epsilon} = (\epsilon \delta + 1)/\tau$.

The SOC for the 0D model is given as:
\begin{equation}\label{eq:soc_t}
	\text{SOC}(t) = \text{SOC}^0 - \frac{I}{\bar{c}_{V} V_e F \epsilon \tilde{ \epsilon} } \left(\frac{\epsilon \delta + e^{-\tilde{ \epsilon} t}}{1+ \epsilon \delta} - 1 - \frac{\epsilon \delta}{\tau} t \right)
\end{equation}
where $\bar{c}_{V}$ is the total vanadium concentration of a half-cell, and $\text{SOC}^0$ is the initial SOC value. Consequently, the concentration solutions in Table \ref{table:conc_soln} can be expressed as functions of SOC, i.e., $c_i(t) = c_i (\text{SOC}(t))$ \cite{he2021physics}.

The reaction kinetics associated with the 0D VRFB model follows the expressions given in the literature \cite{Shah2011,Sharma2015,Eapen2019,he2021physics}.
In the following exposition, we adopt the subscripts "$n$" and "$p$" for the quantities associated with the \textit{negative} and \textit{positive} electrodes for brevity.
With given species concentration $c_i$ and a given applied current $I$, the cell voltage $E^{cell}$ consists of three components:
\begin{equation}\label{eq:cell}
	E^{cell} = E^{OCV} + \eta^{act} + \eta^{ohm}
\end{equation}
Here, voltage loss due to the concentration polarization is ignored. In Eq. \eqref{eq:cell}, $E^{OCV}$ is the reversible open circuit voltage (OCV) that can be approximated by the Nernst equation \cite{Knehr2011,Eapen2019},
\begin{equation}\label{eq:kin_rev}
	E^{OCV} = E_p^0 - E_n^0 
	+ \frac{RT}{F} \ln \left( \frac{c_{\text{V(II)}} c_{\text{V(V)}} c_{\text{H}^+_p} c^2_{\text{H}^+_p}}
	{c_{\text{V(III)}} c_{\text{V(IV)}}   c_{\text{H}^+_n}  c_{\text{H}_2 \text{O}_p} } \right)
\end{equation}
where $E_n^0$ and $E_p^0$ are the negative and positive equilibrium potentials, respectively. 
The $\eta^{act}$ is the activation overpotential described by the Bulter--Volmer equation~\cite{newman2012electrochemical},
\begin{equation}\label{eq:eta_act}
	\eta^{act} = \eta_{p} - \eta_{n}
\end{equation}
\begin{equation*}\label{eq:eta_neg}
	\eta_{n} = -\frac{2RT}{ F} \arcsinh \left( \frac{j}{2Fk_n \sqrt{c_{\text{V(II)}} c_{\text{V(III)}} }} \right) 
\end{equation*}
\begin{equation*}\label{eq:eta_pos}
	\eta_{p} = \frac{2RT}{ F} \arcsinh \left( \frac{j}{2Fk_p \sqrt{c_{\text{V(IV)}} c_{\text{V(V)}} }} \right)
\end{equation*}
and $\eta^{ohm}$ represents the sum of ohmic losses in the current collector (e), membrane (m), and electrolyte (e),
\begin{equation}\label{eq:ohm}
	\eta^{ohm} = \left(\frac{2w_c}{\sigma_c} + \frac{w_m}{\sigma_{m}} + \frac{2w_e}{\sigma_e^{eff}}\right) j_{app}
\end{equation}
where the effective conductivity of the porous electrode is expressed as $\sigma_e^{eff} = (1-\epsilon)^{3/2} \sigma_s$ following the Bruggeman correction~\cite{Bird2002}. 
In the above equations, $j$ and $j_{app}$ are the local and nominal current densities, respectively. Other unstated parameter symbols are referred to in Tables \ref{table:exp_model_para} and \ref{table:PNNL_exp}.
\section{Experimental datasets}\label{sec:exp_data}
\subsection{Experiment conditions}
\begin{table}[htb]
	\centering
	\small{
		\caption{The baseline model parameters for the simulation of the VRFB cell.}
		\label{table:exp_model_para}
		\scalebox{0.9}{
			\begin{tabular}{cccc}
				\toprule
				Symbol & Description  & Unit   &  Values   \\
				\hline
				$E_{p}^0$     & Standard equilibrium potential (Positive)   & $\text{V}$    & $1.004$   \\		
				$E_{n}^0$     & Standard equilibrium potential (Negative)   & $\text{V}$    & $-0.26$   \\	
				$n_{d}$     & Drag coefficient  & -    & $2.5$   \\	
				$k_p$      & Standard rate constant at $303$ K (Positive)   & $\text{m}$ $\text{s}^{-1}$   & $1.0 \times 10^{-7}$    \\	
				$k_n$      & Standard rate constant at $303$ K (Negative) & $\text{m}$ $\text{s}^{-1}$   & $5.0 \times 10^{-8}$   \\
				$S$     & Specific surface area & $\text{m}^{-1}$  & $3.48 \times 10^4$ \\	
				$\epsilon$     & Porosity & - & $0.67$ \\
				$\sigma_{e}$     & Electrode conductivity & $\text{S}$ $\text{m}^{-1}$ & $500$ \\	
				$\sigma_{c}$     & Current collector conductivity & $\text{S}$ $\text{m}^{-1}$ & $9.1 \times 10^{4}$ \\
				$T_{ref}$     & Reference temperature & $K$ & $293$ \\	
				\hline
				$A_e$     & Electrode area  & $\text{m}^{2}$   & $0.002$   \\
				$h_e$     & Electrode length  & $\text{m}$   & $0.05$   \\	
				$b_e$     & Electrode thickness  & $\text{m}$   & $0.04$   \\
				$w_e$     & Electrode width  & $\text{m}$   & $0.004$   \\
				$w_c$   & Current collector width  & $\text{m}$   & $0.015$    \\ 
				$w_m$    & Membrane width  & $\text{m}$   & Ref. to Table \ref{table:PNNL_exp}    \\
				$V_r$     & Reservoir volume  & $\text{m}^{3}$   & Ref. to Table \ref{table:PNNL_exp}     \\			
				\bottomrule
			\end{tabular}%
		}
	}
\end{table}
\begin{table}[htb]
	\centering
	\caption{Operating conditions of 12 experiments for a single-cell structure (the experimental data are the same as \cite{he2021physics}). The membrane width is $w_m = 1.27 \times 10^{-2}$ for Nafion 115 and $w_m = 5.08 \times 10^{-3}$ cm for Nafion 212. All experiments are performed at room temperature ($T=298 K$).}
	\label{table:PNNL_exp}
	\scalebox{0.68}{
		\begin{tabular}{cccccccccc}
			\toprule
			Case  &  $c^0_{V}$  & $c_{\text{H}^+_p}^0$  &  	$ c_{\text{H}^+_n}^0$ &  $ c_{\text{H}_2 \text{O},p}$  & $ c_{\text{H}_2 \text{O},n}$ & $ \omega$  &  I  & $V_r$  &  Membrane \\
			ID			  & [mol $\text{m}^{-3}$]  & [mol $\text{m}^{-3}$]  & [mol $\text{m}^{-3}$]  &  [mol $\text{m}^{-3}$]  &  [mol $\text{m}^{-3}$]    &  [$\text{ml}$ $ \text{min}^{-1}$]  & [$ \text{A}$ ]   & [$\text{m}^{3}$] &  \\
			\hline
			1 & $1.5 \times 10^{3}$  & $3.85 \times 10^{3}$  & $3.03 \times 10^{3}$ & $4.46 \times 10^{4}$ & $4.61 \times 10^{4}$ & 30 & 0.5   & $2 \times 10^{-5}$ &  Nafion 115\\ 
			2 & $1.5 \times 10^{3}$  & $3.85 \times 10^{3}$  & $3.03 \times 10^{3}$ & $4.46 \times 10^{4}$ & $4.61 \times 10^{4}$ & 20 & 0.75   & $8 \times 10^{-5}$ &  Nafion 115\\ 
			3 & $2 \times 10^{3}$  & $5 \times 10^{3}$  & $3 \times 10^{3}$ & $4.75 \times 10^{4}$ & $4.95 \times 10^{4}$ & 20 & 0.5   & $5 \times 10^{-5}$ &  Nafion 115\\ 
			4 & $2 \times 10^{3}$  & $5 \times 10^{3}$  & $3 \times 10^{3}$ & $4.75 \times 10^{4}$ & $4.95 \times 10^{4}$ & 20 & 0.69  & $4.5 \times 10^{-5}$ &  Nafion 115\\ 
			5 & $2 \times 10^{3}$  & $5 \times 10^{3}$  & $3 \times 10^{3}$ & $4.75 \times 10^{4}$ & $4.95 \times 10^{4}$ & 20 & 0.75  & $4.5 \times 10^{-5}$ &  Nafion 115\\ 
			6 & $2 \times 10^{3}$  & $5 \times 10^{3}$  & $3 \times 10^{3}$ & $4.75 \times 10^{4}$ & $4.95 \times 10^{4}$ & 20 & 1.5   & $4.5 \times 10^{-5}$ &  Nafion 115\\ 
			7 & $2 \times 10^{3}$  & $5 \times 10^{3}$  & $3 \times 10^{3}$ & $4.75 \times 10^{4}$ & $4.95 \times 10^{4}$ & 20 & 0.5   & $5 \times 10^{-5}$ &  Nafion 212\\ 
			8 & $2 \times 10^{3}$  & $5 \times 10^{3}$  & $3 \times 10^{3}$ & $4.75 \times 10^{4}$ & $4.95 \times 10^{4}$ & 20 & 0.4   & $3 \times 10^{-5}$ &  Nafion 212\\ 
			9 & $2 \times 10^{3}$  & $5 \times 10^{3}$  & $3 \times 10^{3}$ & $4.75 \times 10^{4}$ & $4.95 \times 10^{4}$ & 20 & 0.4   & $2.5 \times 10^{-5}$ &  Nafion 212\\ 
			10 & $2 \times 10^{3}$  & $5 \times 10^{3}$  & $3 \times 10^{3}$ & $4.75 \times 10^{4}$ & $4.95 \times 10^{4}$ & 20 & 0.5   & $4\times 10^{-5}$ &  Nafion 212\\ 
			11 & $2 \times 10^{3}$  & $5 \times 10^{3}$  & $3 \times 10^{3}$ & $4.75 \times 10^{4}$ & $4.95 \times 10^{4}$ & 20 & 1.0   & $2\times 10^{-5}$ &  Nafion 212\\ 
			12 & $1.5 \times 10^{3}$  & $3.85 \times 10^{3}$  & $3.03 \times 10^{3}$ & $4.46 \times 10^{4}$ & $4.61 \times 10^{4}$ & 20 & 0.4   & $3\times 10^{-5}$ &  Nafion 212\\ 
			\bottomrule
		\end{tabular}
	}
\end{table}

The high-fidelity experimental VRFB data are collected at Pacific Northwest National Laboratory for training the proposed ePCDNN \cite{he2021physics,Bao2019}. 
As shown in Table \ref{table:PNNL_exp}, 12 experiments were performed on a single flow-through cell with varied operating conditions (total species concentrations, flow rate, and applied current) and cell components (electrolyte tank volume and membrane type). 
The electrode area ($A_e = 20$ $\text{cm}^2$), electrode thicknesses ($w_e = 0.4$), and current collector thickness ($w_c = 1.5$ $\text{cm}$) are kept the same for each experiment.
The Nafion 115 membrane is used for experiments 1--6 with a thickness of $w_m = 1.27 \times 10^{-2}$ cm. The Nafion 212 membrane is used for experiments 7--12 with a thickness of $w_m = 5.08 \times 10^{-3}$ cm.
The model parameters that are used for the 0D model as the baseline results are given in Table \ref{table:exp_model_para}. Herein, the parameters
$S$, $k_n$, $k_p$, and $\sigma_{e}$ will be estimated using the experimental data. 

\begin{figure}[htb!]
	\centering
	\subfloat[] {\includegraphics[width=2.5in]{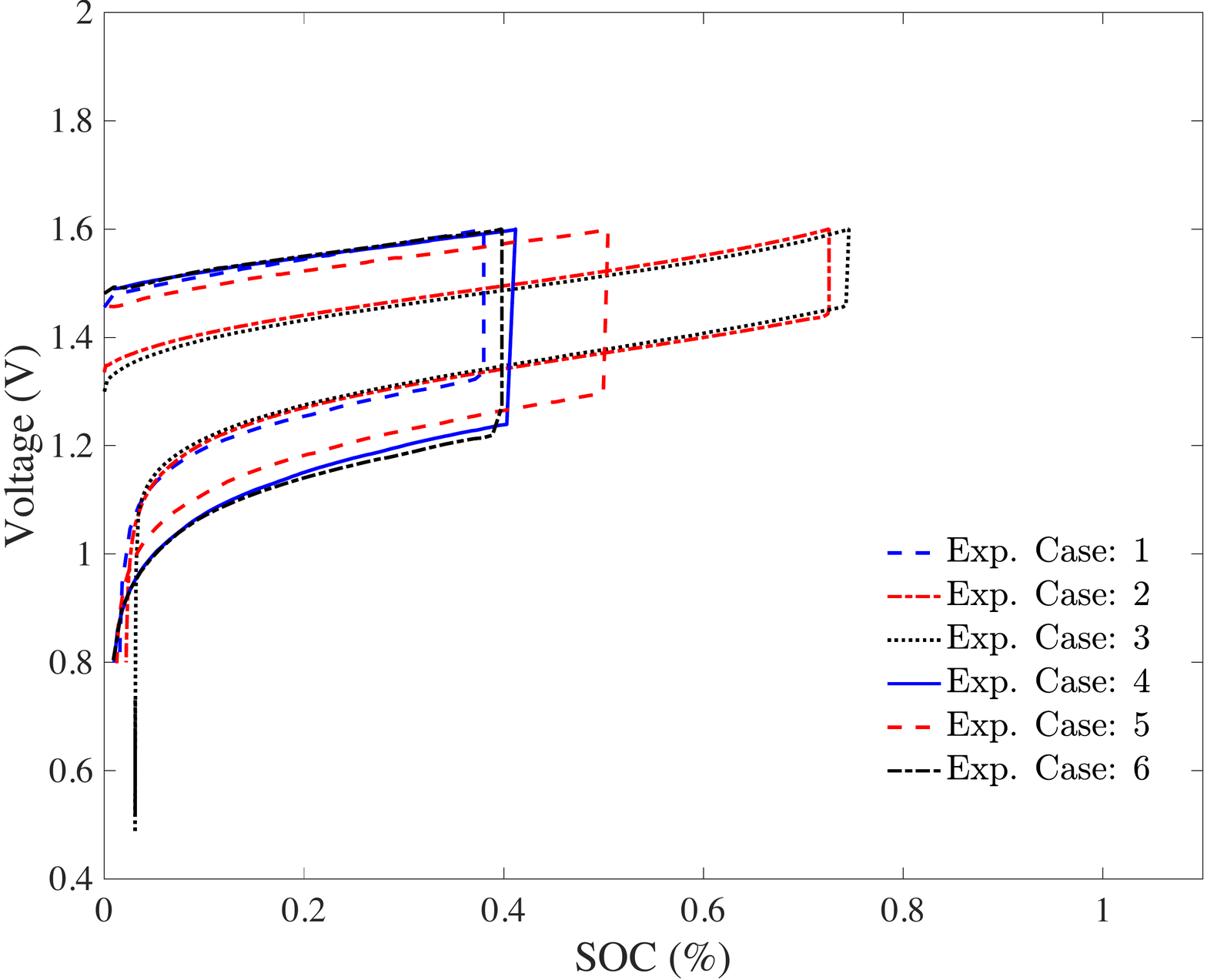}}
	\subfloat[] {\includegraphics[width=2.5in]{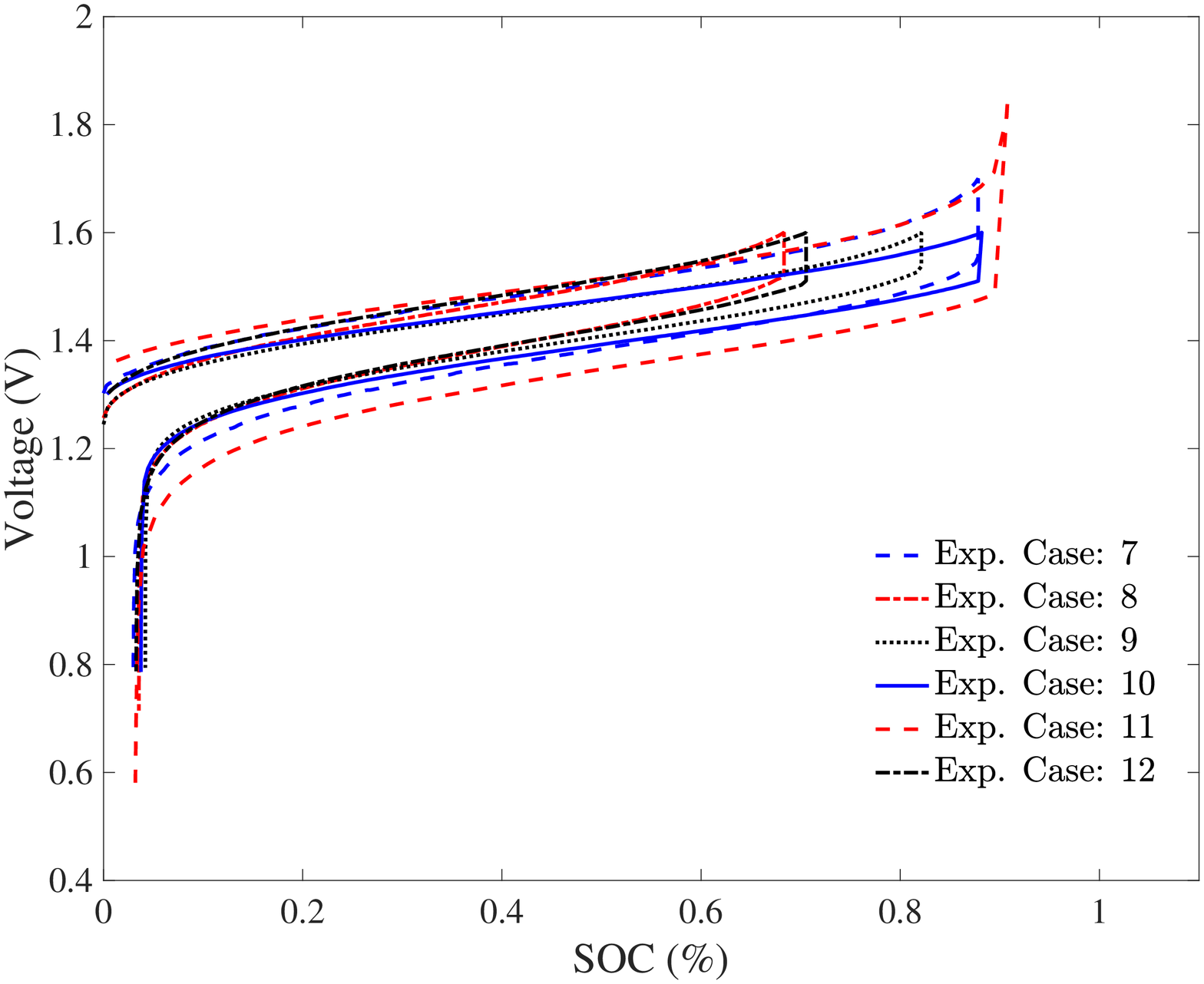}} \\
	\caption{The SOV-Voltage charge--discharge curves (third cycle) for the 12 PNNL experiments given in (\cite{he2021physics} , Table \ref{table:PNNL_exp}).}
	\label{fig:exp_raw_data}
\end{figure}

For each experiment, the charge--discharge data are recorded with multiple cycles, and a good coulombic efficiency is observed starting from the third charge--discharge cycle. Therefore, the voltage responses measured from the third cycle are selected as the experimental data for testing the ePCDNN model.
An overview of the 12 experiment SOV-V curves is given in Fig. \ref{fig:exp_raw_data}.  The SOC value is converted from the corresponding time step using the SOC-t relation given by Eq. \eqref{eq:soc_t}.

\subsection{Parameter selection for ePCDNN}
Consistent with PCDNN, operating conditions including average electrolyte flow velocity $\tilde{u}$, applied current $I$, and initial vanadium concentration $c^0_V$ are selected in the DNN models here as the input vector $\vec{x}=\{\tilde{u}, I, c^0_V\}$ to represent different experimental operation conditions. Note that the evolution of concentration and voltage during the charge--discharge cycles also depends on the selection of the initial conditions.

For the 0D model in Section \ref{sec:model_0D}, the specific area for reaction $S$, the reaction rate constants $k_n$ and $k_p$, and the electrode conductivity $\sigma_e$ are chosen as the identifiable model parameters.
These selected parameters are encoded in a parameter vector $\vec{\mu} = (S, k_n, k_p, \sigma_e)$ of $m=4$ dimensions, and their baseline values are provided in Table \ref{table:exp_model_para}.
In the following sections, we assume that the selected parameters are unknown functions of the operating conditions, i.e., $\vec{\mu} = \vec{\mu}(\vec{x})$. One objective of this study is learning these parameter functions from a given set of experimental data and providing accurate voltage prediction for various operating conditions.

\section{Methods}\label{sec:Method}

\subsection{Enhanced physics-constrained deep neural networks}\label{sec:ePCDNN}
\begin{figure}[ht!]
	\centering
	\includegraphics[angle=0,width=5in]{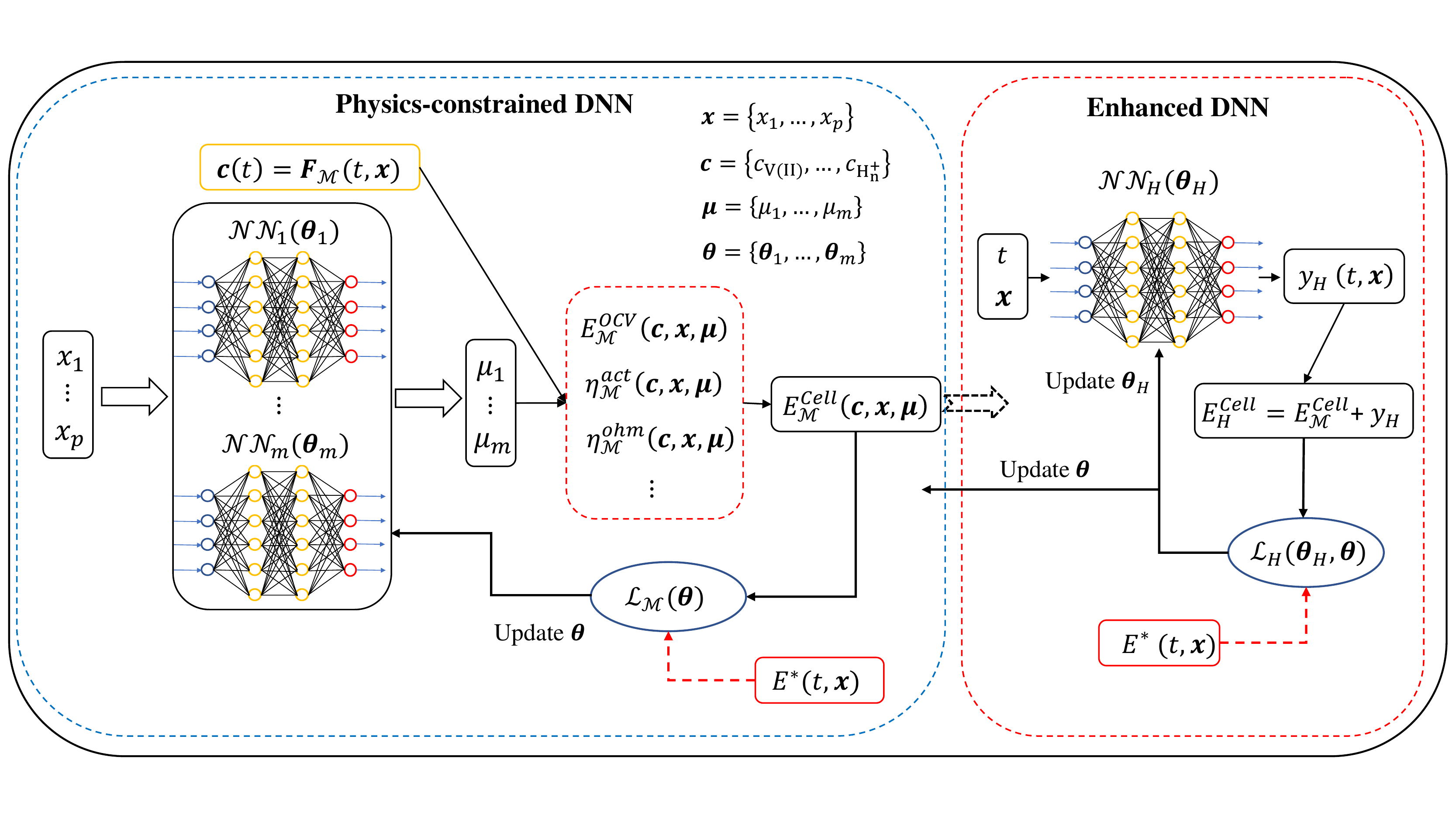}
	\caption{
	Schematic of the proposed ePCDNN model that consists of two network components: (1) the PCDNN \cite{he2021physics} (in the blue dashed box) encoded with the 0D electrochemical model and (2) the enhanced DNN (in the red dashed box).
	In the PCDNN component, $m$ individual neural networks are used to relate the operating conditions $\{x_i\}_{i=1}^p$ and the $m$ model parameters $\{\mu_i\}_{i=1}^m$. The enhanced DNN is used to learn a high-fidelity correction $y_H$ that is superposed to the PCDNN voltage prediction $E_{\mathcal{M}}^{Cell}$. The voltage prediction of ePCDNN is $E_{\mathcal{H}}^{Cell} = E_{\mathcal{M}}^{Cell}  + y_H$.
	$\mathcal{M}$ denotes a predefined physical model used to compute the concentrations of species $\vec{c}$ and cell voltages.
	}
%
	\label{fig:sche_mpinn}
\end{figure}

In this section, the ePCDNN approach is developed based on the PCDNN framework \cite{he2021physics}. In addition to retaining efficient parameter identification, the new approach improves the predictive capacity of cell voltage by introducing an additional enhanced DNN component. The ePCDNN model is depicted in Fig. \ref{fig:sche_mpinn} by coupling the standard PCDNN model (enclosed by the blue dashed box) with the enhanced DNN (in the red dashed box).

In the PCDNN component, each model parameter $\mu_i$ is approximated by a DNN (the fully connected feed-forwad DNN is adopted here), 
\begin{equation}\label{eq:mu_dnn}
	\hat{\mu}_i (\vec{x}; \vec{\theta}_i)  = \mathcal{NN}_i(\vec{x};\vec{\theta}_i),  \quad \text{for} \;  i = 1,...,m
\end{equation}
where $\mathcal{NN}_i$ denotes a DNN model with the trainable weight coefficients $\vec{\theta}_i$ and where $\vec{x}$ is the input vector composed of experimental operating conditions of interest. 
The collection of ${\{\theta_i\}}_{i=1}^m$ is denoted as $\vec{\theta}$. Then, the cell voltage prediction is given as
\begin{equation}\label{eq:cell_pinn}
	E^{cell}_{\mathcal{M}}(\vec{c}, \vec{x}, \hat{\vec{\mu}}(\vec{\theta})) = E^{OCV}_{\mathcal{M}}(\vec{c}, \vec{x}, \hat{\vec{\mu}}(\vec{\theta})) + \eta^{act}_{\mathcal{M}}(\vec{c}, \vec{x}, \hat{\vec{\mu}}(\vec{\theta})) + \eta^{ohm}_{\mathcal{M}}(\vec{c}, \vec{x}, \hat{\vec{\mu}}(\vec{\theta}))
\end{equation}

The enhanced DNN is expressed as
\begin{equation}
	y_H (t, \vec{x}; \vec{\theta}_H) = \mathcal{NN}_H(t, \vec{x}; \vec{\theta}_H)
\end{equation}
where $\vec{\theta}_H$ is the network parameters.

Considering the SOC--time relation derived in Eq. \eqref{eq:soc_t}, we can estimate the $\text{SOC}(t)$ value for any given time $t$. We choose to directly use SOC instead of $t$ as the direct network input of the enhanced DNN $\mathcal{NN}_H$ because the SOC value is normalized within $[0,1]$ and is suitable for network training. However, we still denote it as $\mathcal{NN}_H (\text{SOC}(t), \vec{x}; \vec{\theta}_H):= \mathcal{NN}_H (t, \vec{x}; \vec{\theta}_H)$ for brevity.

In the ePCDNN approach, we define the following loss function:
\begin{equation}\label{eq:loss_epinn}
		\mathcal{L} (\vec{\theta}, \vec{\theta}_H)  = \lambda \mathcal{L}_{\mathcal{M}} (\vec{\theta}) + (1-\lambda)\mathcal{L}_{H} (\vec{\theta}_H, \vec{\theta})
\end{equation}
where $\mathcal{L}_{\mathcal{M}} (\vec{\theta}) $ measures the mean square error between the PCDNN prediction and the given measurements $E^*$: 
\begin{equation}\label{eq:loss_pinn}
	\mathcal{L}_{\mathcal{M}} (\vec{\theta}) = \frac{1}{N^{x}} \sum_{q=1}^{N^{x}} \frac{1}{N^t_q} \sum_{l=1}^{N^t_q} [E^{cell}_{\mathcal{M}}(t_l, \vec{x}_q; \vec{\theta}) - E^*(t_l,\vec{x}_q)]^2
\end{equation}
and $\mathcal{L}_{H} (\vec{\theta}_H,\vec{\theta})$ denotes the error of the enhanced voltage prediction $E^{cell}_H = E^{cell}_{\mathcal{M}} + y_H$ against the measurements $E^*$: 
\begin{equation}\label{eq:loss_2}
	\mathcal{L}_{H} (\vec{\theta}_H,\vec{\theta}) = \frac{1}{N^{x}} \sum_{q=1}^{N^{x}} \frac{1}{N^t_q} \sum_{l=1}^{N^t_q} [E^{cell}_H (t_l, \vec{x}_q; \vec{\theta}, \vec{\theta}_H) - E^*(t_l,\vec{x}_q)]^2,
\end{equation}
where $N^x$ is the number of experimental datasets, $N^t_q$ is the number of measurements collected from each experiment, and $t_l$ ($l=1,...,N_q^t$) denotes the corresponding times for measurements. The experimental measurements are as shown in Fig. \ref{fig:exp_raw_data}.
The DNN parameters $\vec{\theta}$ and $\vec{\theta}_H$ are trained by minimizing the above loss function $\mathcal{L} (\vec{\theta}, \vec{\theta}_H)$. The L-BFGS-B \cite{Byrd1995} and Adam \cite{Kingma2015} methods are selected as the gradient descent optimization algorithms. At the beginning of each training process, the DNN weights are randomly initialized using the Xavier scheme \cite{Glorot2010}.

The $\lambda$ in Eq. \ref{eq:loss_epinn} defines the weight of contributions from the loss function of $\mathcal{L}_{\mathcal{M}} $ and $\mathcal{L}_{\mathcal{H}} $. Minimizing the loss term $\mathcal{L}_{\mathcal{M}}$ optimizes the model parameter functions $\vec{\mu}(\vec{x};\vec{\theta})$ while the minimization of the loss term $\mathcal{L}_{H}$ optimizes the voltage prediction $E^{cell}_H$ with respect to the given measurements. Also, if the $\mathcal{L}_{H}$ term is neglected during training procedures and assuming $y_H=0$, the standard PCDNN solution \cite{he2021physics} is recovered.

We note that the trained PCDNN component $E^{cell}_{\mathcal{M}}$ in the ePCDNN and PCDNN methods is different because different loss functions are considered. Our numerical results show that the minimization of $\mathcal{L}_{H} (\vec{\theta}_H,\vec{\theta})$ in ePCDNN can also improve the prediction of $E^{cell}_{\mathcal{M}}.$ In this work, we demonstrate that the optimal performance is achieved for $\lambda$ values in the range of $[0.25,0.5]$.  The $\lambda = 0.5$ is selected for the following test by default. Further details of the loss function design will be discussed in Section~\ref{sec:lossfunctiondesign}.

\section{Results and discussion}\label{sec:Results}
\subsection{Parameter estimation and voltage prediction}\label{sec:res_exp_case1}
\begin{figure}[htb!]
	\centering
	\includegraphics[angle=0,width=3in]{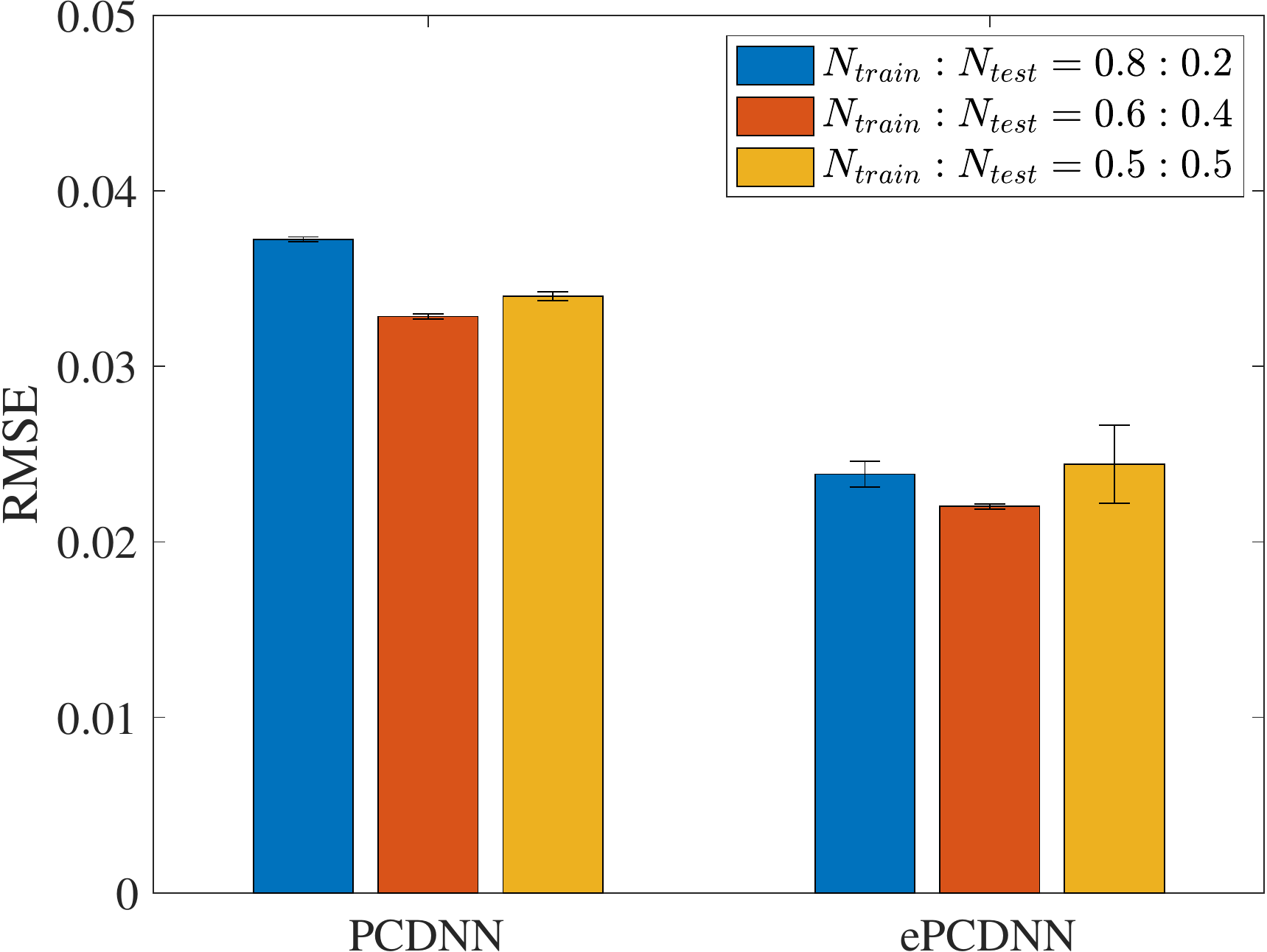}
	\caption{The comparison of the RMSEs in the predicted cell voltage for the ePCDNN and PCDNN models using different sizes of training data. The bars correspond to one standard deviation of the RMSEs and quantify uncertainty due to random initialization of DNNs.}
	\label{fig:comp_column}
\end{figure}

\begin{figure}[htb!]
	\centering
	\includegraphics[angle=0,width=5in]{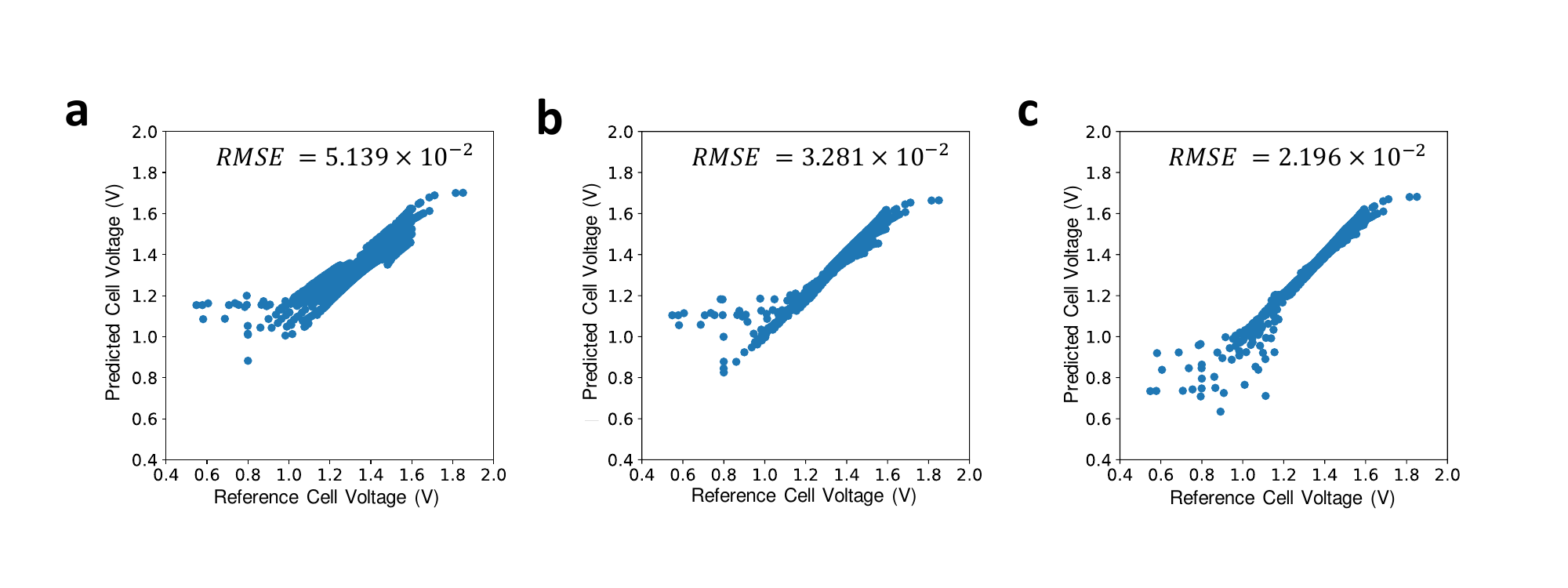}
	\caption{
	The parity plots to compare the ground-truth voltage data with (a) the baseline 0D model, (b) the trained PCDNN model, and (c) the ePCDNN model. The DNN structure $4 \times 30$ is used for the enhanced DNN.}
	\label{fig:comp_r2_plot}
\end{figure}

We use the 12 experimental cases shown in Fig. \ref{fig:exp_raw_data} to demonstrate the enhanced performance of the proposed ePCDNN approach over PCDNN and the 0D model, especially for the regions where the voltage changes fast.

The experimental data split percentage for training and testing are set to $40\%:60\%$, $60\%:40\%$, and $80\%:20\%$ for both the ePCDNN and PCDNN models. Model prediction accuracy is measured on the testing data with the root-mean-square error (RMSE) and is shown in Fig. \ref{fig:comp_column}. 
The statistics of the RMSE are computed from five independent trainings where the DNNs are randomly initialized.
Fig. \ref{fig:comp_column} shows that the proposed ePCDNN model reduces the test errors by about 30\% compared to the PCDNN approach regardless of the size of the training dataset used. 

The parity plots that compare the predicted voltage and the ground-truth experimental data are given in Fig. \ref{fig:comp_r2_plot}. For the voltage $E^*$> 1.2 V, the data points are close to the center line $y=x$, indicating that the 0D model, PCDNN, and ePCDNN provide reasonable predictions in those regions. For the voltage $E^*$< 1.2 V, the 0D model starts to deviate from the center line, and most of the predictions overestimate the voltages. The PCDNN performs better with more data points close to the $y=x$ compared to the 0D model at $E^*$< 1.2 V.  Among the three models, the ePCDNN achieves the best accuracy at both the high- and low-voltage regions.  This indicates that the enhanced DNN part did a good job correcting the missing physics from the 0D numerical model at the low-voltage region.


\begin{figure}[htb!]
	\centering
	\includegraphics[angle=0,width=6in]{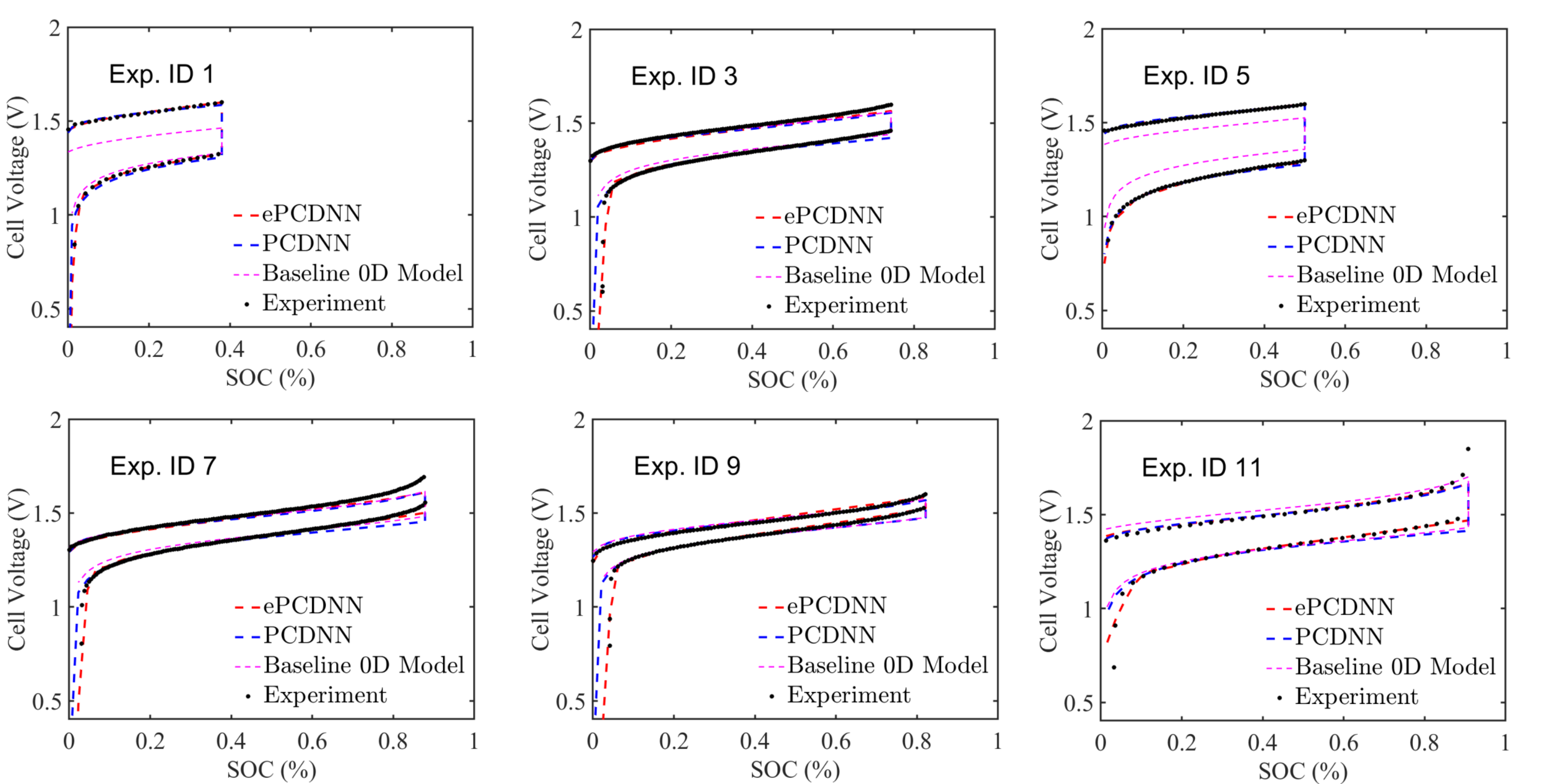}
	\caption{Comparison of the ePCDNN model, 0D model (baseline), and ePCDNN performance with the six selected charge--discharge curves in the VRFB experiments.}
	\label{fig:plot_N12_p3s}
\end{figure}

To visualize the improvement of ePCDNN, the charge--discharge curves from six experiments were selected for comparison with PCDNN and the baseline 0D model, as shown in Fig. \ref{fig:plot_N12_p3s}. The results show that the ePCDNN achieves the best agreement with experimental data.  The discharge curve tails are captured by the ePCDNN, while the 0D model and PCDNN can overestimate the voltages for several experiments.

\subsection{Loss function design}\label{sec:lossfunctiondesign}
We explore the effect of key ePCDNN hyperparameters on voltage prediction performance.  All other parameters are held constant in the hyperparameter study.  The structure of the enhanced DNN used in the ePCDNN is investigated by varying the sizes of $n_l $ and $m_l$, as shown in Table \ref{table:Effect_ENN Size}. The parameter $n_l$ stands for the number of hidden layers, and $m_l$ is the neurons in each layer.  We show the RMSE, the $L^{\infty}$ error, and the computational time for each enhanced DNN structure configuration.  Overall, a lower RMSE is obtained with a neural network of larger size.  However, the computational time increases as the number of neurons is increased because the training becomes slower. 

When the enhanced DNN has four hidden layers, the RMSE drops to below $2\times10^{-2}$. Further increasing the network size only marginally improves the RMSE.  In terms of the $L^{\infty}$ error, the $2\times20$ or $4\times40$ configuration results in the best accuracy. Considering the performance and efficiency of the networks, the $4\times40$ size is selected as the optimal setup for the enhanced DNN.

\begin{table}[htb]
	\centering
	\caption{The effect of the size of the enhanced DNN size on voltage prediction}
	\label{table:Effect_ENN Size}
	\scalebox{0.8}{
		\begin{tabular}{cccc}
			\toprule
			Size & RMSE & $L^{\infty}$ Error & Time \\
			\hline
			$1 \times 20$ & $2.88\times 10^{-2}$ & $6.32 \times 10^{-1}$ &75\\
			$2 \times 20$ & $2.14\times 10^{-2}$ & $3.84 \times 10^{-1}$ &157\\
			$2 \times 30$ & $2.16\times 10^{-2}$ & $5.55 \times 10^{-1}$ &189\\
			$3 \times 30$ & $2.23\times 10^{-2}$ & $5.55 \times 10^{-1}$ &144\\
			$3 \times 40$ & $2.27\times 10^{-2}$ & $4.05 \times 10^{-1}$ &180\\
        	$4 \times 40$ & $1.95\times 10^{-2}$ & $3.98 \times 10^{-1}$ &221\\
			$4 \times 60$ & $1.98\times 10^{-2}$ & $4.00 \times 10^{-1}$ &329\\	 
	        $5 \times 40$ & $1.93\times 10^{-2}$ & $5.55 \times 10^{-1}$ & 305\\
        	$6 \times 40$ & $1.95\times 10^{-2}$ & $4.18 \times 10^{-1}$ & 367\\
			\bottomrule
		\end{tabular}
	}
\end{table}

The construction of the loss function is important to reduce the voltage estimation error. The loss function combines $\mathcal{L}_{\mathcal{M}}$ and $\mathcal{L}_{\mathcal{H}}$ with a weight coefficient denoted by $\lambda.$  To determine the optimal $\lambda$ value, a parameter study was carried out by varying $\lambda$ and the train/testing data ratio. 

When $\lambda$ approaches 1, the loss function $	\mathcal{L} (\vec{\theta}, \vec{\theta}_H) $ approaches that for the PCDNN, with no additional corrections to account for the tail voltage discrepancy.  As shown in Table \ref{table:Effect_Loss}, the RMSE and $L^{\infty}$ error are both the largest. With $\lambda$ close to 0, training the neural network updates the weights and biases of the physics-constrained DNN. However, there is no separate enforcing of the physics-constrained DNN predictions to conform to the measurements. Therefore, the output $y_H$ of the second DNN is no longer necessarily a correction to $E_{\cal{M}}^{cell}$. This makes it more difficult for the neural network to predict unseen conditions when there is a large dataset split (when $N_{train}:N_{test}$ is closer to 0.5:0.5). The $L^{\infty}$ error increases, which indicates that the ePCDNN can have reduced accuracy in predicting the end of the discharge curve, where the slope is the largest.  Therefore, it is reasonable to select $\lambda$ in the $[0.25,0.5]$ range, which balances the enforced physics and the enhanced DNN.

\begin{table}[htb]
	\centering
	\caption{The effect of loss function design on voltage prediction}
	\label{table:Effect_Loss}
	\scalebox{0.8}{
		\begin{tabular}{ccccccc}
			\toprule
			$\lambda$ & \multicolumn{2}{c}{$N_{train}$:$N_{test}$ = 0.8:0.2}  &  \multicolumn{2}{c}{$N_{train}$:$N_{test}$ = 0.6:0.4} &\multicolumn{2}{c}{$N_{train}$:$N_{test}$ = 0.5:0.5}\\
			 & RMSE & $L^{\infty}$ Error & RMSE & $L^{\infty}$ Error & RMSE & $L^{\infty}$ Error \\
			\hline
			 0 & $1.86\times 10^{-2}$ & $3.69\times 10^{-1}$ & $1.92\times 10^{-2}$ & $4.03\times 10^{-1}$ &  $1.98\times 10^{-2}$ & $4.50 \times 10^{-1}$\\
			 0.25 & $1.87\times 10^{-2}$ & $3.68\times 10^{-1}$ & $1.92\times 10^{-2}$ & $3.98\times 10^{-1}$ &  $1.97\times 10^{-2}$ & $4.42 \times 10^{-1}$\\
			 0.5 & $2.02\times 10^{-2}$ & $3.91\times 10^{-1}$ & $2.04\times 10^{-2}$ & $3.98\times 10^{-1}$ &  $2.03\times 10^{-2}$ & $4.20 \times 10^{-1}$\\
			 0.75 &  $2.07\times 10^{-2}$ & $3.95\times 10^{-1}$ & $2.10\times 10^{-2}$ & $3.90\times 10^{-1}$ &  $2.17\times 10^{-2}$ & $4.03 \times 10^{-1}$\\
	          1.0 &  $3.46\times 10^{-2}$ & $6.36\times 10^{-1}$ & $3.46\times 10^{-2}$ & $6.36\times 10^{-1}$ &  $2.46\times 10^{-2}$ & $6.35 \times 10^{-1}$\\
			\bottomrule
		\end{tabular}}
\end{table}

\subsection{The effect of data sampling}
\begin{figure}[htb!]
	\centering
	\includegraphics[angle=0,width=4.5in]{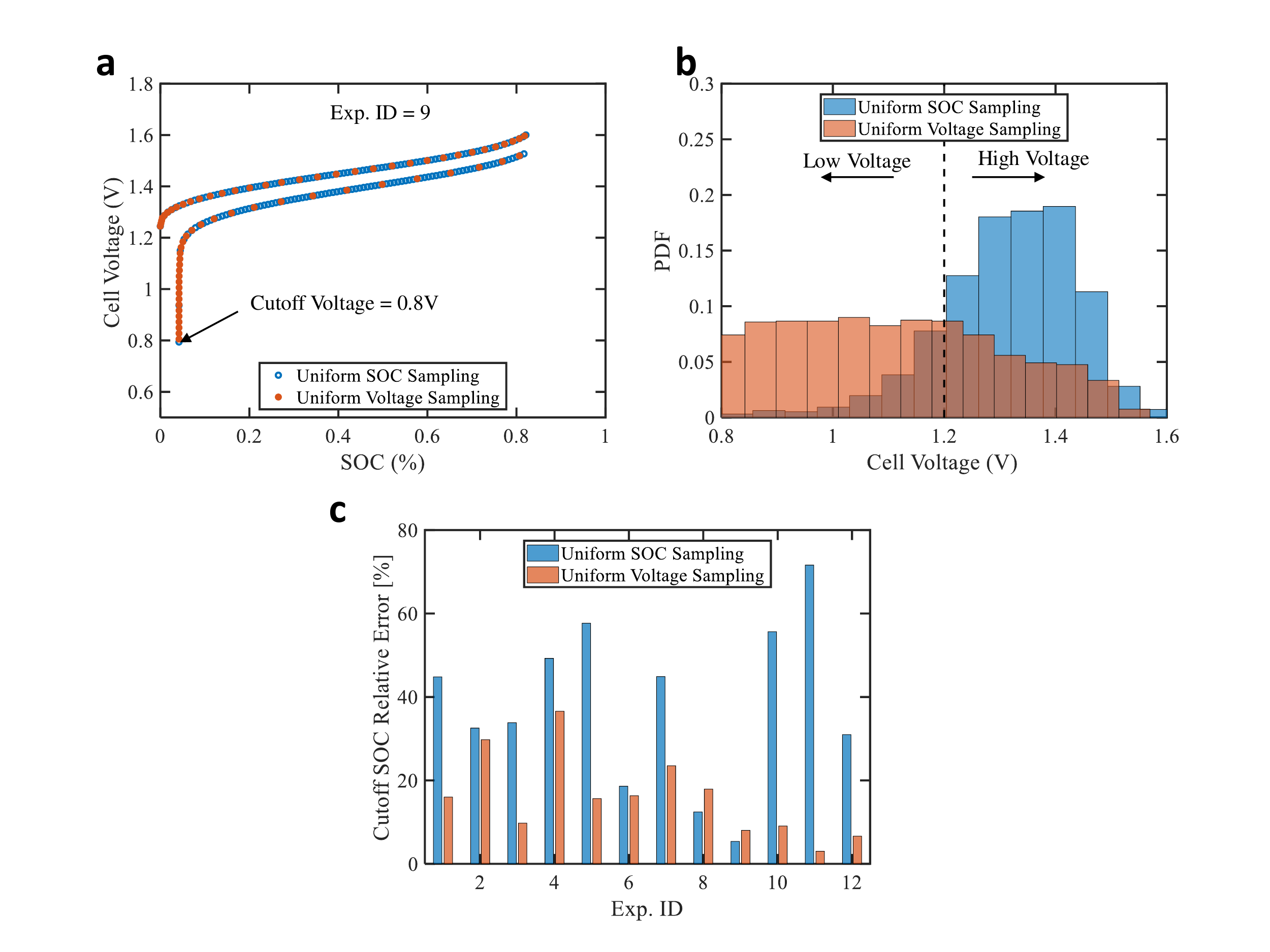}
	\caption{(a) Discharge curve of Exp. ID = 9 with a uniform SOC and voltage sampling method, (b) the voltage data point distribution, and (c) the resultant cutoff time prediction accuracy with two different sampling methods. }
	\label{fig:SamplingEffect}
\end{figure}

Training a neural network can be challenging with an imbalanced dataset, especially when the target values are continuous\cite{yang2021delving}. In the original experiment measurement, the voltage responses are sampled uniformly in time.  This translates to a nearly uniform SOC distribution for each experimental case, as shown in Fig. \ref{fig:SamplingEffect}(a). We denote this as the uniform SOC sampling method in the following discussion. With this sampling method, the distribution of voltage data points is skewed, with more data points in the high-voltage region (V > 1.2), as shown in Fig. \ref{fig:SamplingEffect}(b).  The entire charge curve and the majority of the discharge curve fall in the high-voltage value regions.  Those curve segments are relatively flat with small slopes.  In contrast, only $17\%$ of the total voltage measurement points are in the low-voltage range (V < 1.2). These data consist of the discharge curve tails, which have steep slopes. 

To investigate the influence of the imbalanced dataset on ePCDNN performance, we instead sample the cell charge/discharge curve uniformly in the voltage. The sampled curve is shown as red dots in Fig. \ref{fig:SamplingEffect}(a). From Fig. \ref{fig:SamplingEffect}(b), we see that the uniform voltage sampling method provides a more balanced voltage data distribution. At the low-voltage region (0.8--1.2V) where the slope is large, the data point density is more uniform, while it decreases more rapidly in the high-voltage region (1.2--1.6V) where the voltage curve slope is small. 

To compare the performance of the two sampling methods, the accuracy in predicting the SOC cutoff is evaluated for each experiment in our dataset.  Recall that the cutoff SOC is defined as the location where the discharge voltage drops below 0.8 V,which is used to protect the battery from excessive discharge. At the end of discharge, the amounts of V(II) and V(V) are nearly completely consumed and the experimentally measured cutoff SOC will be close to zero. The relative error in cutoff SOC prediction is larger compared to the RMSE criteria, which compare the overall error throughout the entire voltage curve. 

As seen in Fig. \ref{fig:SamplingEffect}(c), when using the uniform SOC sampling method, the relative error for the cutoff SOC can reach up to $71\%,$ while the relative error averaged across all the experiments is $38\%$. On the other hand, using the uniform voltage sampling method, the cutoff SOC prediction accuracy improves significantly with an average relative error of $16\%$.  The maximum relative error is measured at $36\%$ in the fourth experiment. This suggests that by considering the features of the studied VRFB voltage curves, a more appropriate sampling strategy can reduce the influence of the imbalanced dataset and improve the ePCDNN prediction performance for the curve tail region.

\subsection{Prediction for unseen experiments}\label{sec:res_exp_case2}
\begin{figure}[htb!]
	\centering
	\includegraphics[angle=0,width=4.5in]{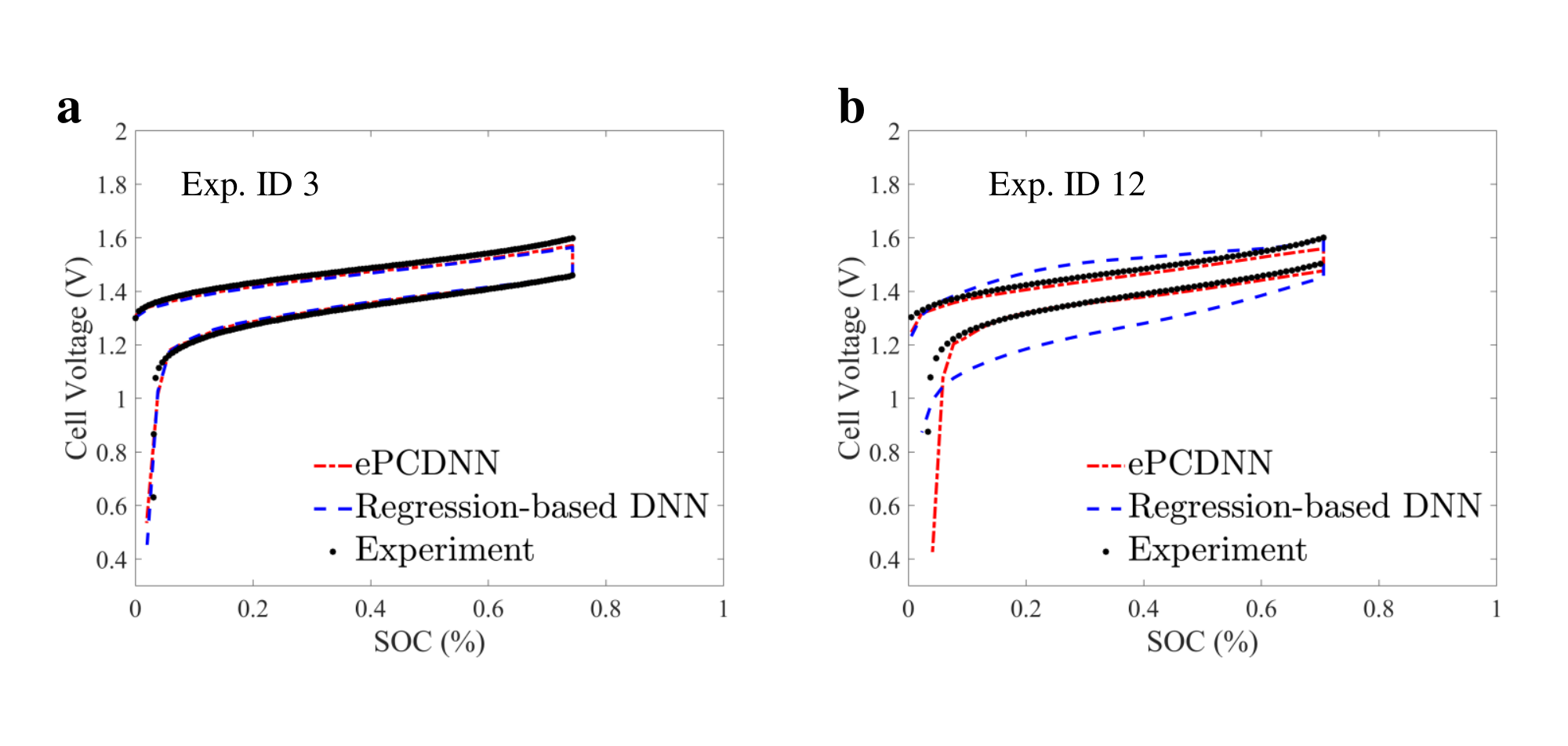}
	\caption{The comparison of the ePCDNN and regression-based DNN model accuracy and generalization for unseen experiments of (a) case ID 3 and (b) case ID 12.  }
	\label{fig:unseen}
\end{figure}

In this section, we carry out a test on the predictive ability of ePCDNN for unseen experiments, i.e., its generalization ability. A regression-based DNN (no physics) and the ePCDNN were trained using only 11 experimental datasets, and the remaining experimental dataset was used for testing. Fig.~\ref{fig:unseen} shows results when experiments 3 and 12 are respectively used as the unseen experiment. The experiment 12 is the most challenging case with input parameter outside the range of the training data.   When experiment 3 is unseen, both regression-based DNN and ePCDNN give an accurate voltage prediction with an RMSE of 0.024 V and 0.022 V, respectively. Also, the tail in the discharge curve at a small SOC is well captured.  However, when experiment 12 is unseen, the data-driven DNN captures neither the tail nor the flat region of the voltage curve and the RMSE is 0.082 V. On the other hand, the enhanced PCDNN more reasonably predicts the SOC-V curve with an RMSE value of 0.048 V. Also, the sharp drop in the tail of the discharge curve is accurately predicted. 

From the operating conditions in Table \ref{table:PNNL_exp}, we see that experiment 3 uses the Nafion 115 membrane with $c_V^0 = 2\times10^3 mol/m^{-3}$ and $I = 0.5 A$.  Those particular concentration and current values are also found in some of the experimental datasets used for training. As a result, the prediction task is close to an interpolation task, and the regression-based DNN performs well. This explains the similar accuracy between the regression-based DNN and the ePCDNN. However, experiment 12 uses the thinner Nafion 212 membrane whose $c_V^0$ and proton concentration values have not been seen before in the experiments used for training.  This forces the regression-based DNN to extrapolate, which is known to be a challenge. On the other hand, the physics-constrained ePCDNN is more capable of predicting the unseen experiment because of its superior generalization ability. 

\section{Conclusions}\label{sec:Conclusions}
We have developed a framework, called ePCDNN, for modeling the VRFB system by combining a physics-constrained DNN to learn the parameters of the 0D RFB model and an enhanced DNN whose output is used to correct the prediction of the 0D model. Due to its simplicity, the 0D model can provide a  prediction with low model computational cost. This makes it suitable for implementation in the ePCDNN framework. Combined with the enhanced DNN, the ePCDNN provides both good accuracy and generalization for VRFB voltage prediction.

Our numerical results show that the proposed ePCDNN framework indeed leads to improved predictions compared to the PCDNN method \cite{he2021physics}. The ePCDNN can capture the sharp drop in the tail of the discharge curve, which is not feasible using only the 0D numerical model. We have allowed the loss function form to be more flexible by assigning possibly different weights on its two terms, one for the physics-constrained DNN and one for the enhanced DNN. For applications with high-fidelity physics models, the weight can be increased to emphasize the physics-constrained DNN contribution. On the other hand, if the physics model is simple, the weight of the loss function term for the enhanced DNN can be increased to address the missing physics in the model and to improve the performance. In addition to the studied VRFB system, the developed ePCDNN framework can be applied to different battery systems by replacing the enforced physics model. This makes the ePCDNN framework suitable for aiding the acceleration of battery system design and optimization.

\section*{Acknowledgements}

This research was supported by the Energy Storage Materials Initiative (ESMI) at Pacific Northwest National Laboratory (PNNL).  PNNL is a multi-program national laboratory operated for the U.S. Department of Energy (DOE) by Battelle Memorial Institute under Contract No. DE-AC05-76RL01830.

\bibliography{references.bib,mybib_PINN.bib}

\begin{thebibliography}{10}
\expandafter\ifx\csname url\endcsname\relax
  \def\url#1{\texttt{#1}}\fi
\expandafter\ifx\csname urlprefix\endcsname\relax\def\urlprefix{URL }\fi
\expandafter\ifx\csname href\endcsname\relax
  \def\href#1#2{#2} \def\path#1{#1}\fi

\bibitem{he2021physics}
Q.~He, P.~Stinis, A.~Tartakovsky, Physics-constrained deep neural network
  method for estimating parameters in a redox flow battery, arXiv preprint
  arXiv:2106.11451 (2021).

\bibitem{chen2021carbon}
J.~M. Chen, Carbon neutrality: Toward a sustainable future, The Innovation
  2~(3) (2021).

\bibitem{soloveichik2015flow}
G.~L. Soloveichik, Flow batteries: current status and trends, Chemical reviews
  115~(20) (2015) 11533--11558.

\bibitem{noack2015chemistry}
J.~Noack, N.~Roznyatovskaya, T.~Herr, P.~Fischer, The chemistry of redox-flow
  batteries, Angewandte Chemie International Edition 54~(34) (2015) 9776--9809.

\bibitem{weber2011redox}
A.~Z. Weber, M.~M. Mench, J.~P. Meyers, P.~N. Ross, J.~T. Gostick, Q.~Liu,
  Redox flow batteries: a review, Journal of applied electrochemistry 41~(10)
  (2011) 1137.

\bibitem{tokuda1998development}
N.~Tokuda, T.~Kumamoto, T.~Shigematsu, H.~Deguchi, T.~Ito, N.~Yoshikawa,
  T.~Hara, Development of a redox flow battery system, SEI Tech. Rev. 50 (2000)
  88.

\bibitem{lopez1992optimization}
M.~Lopez-Atalaya, G.~Codina, J.~Perez, J.~Vazquez, A.~Aldaz, Optimization
  studies on a fe/cr redox flow battery, Journal of power sources 39~(2) (1992)
  147--154.

\bibitem{skyllas2003novel}
M.~Skyllas-Kazacos, Novel vanadium chloride/polyhalide redox flow battery,
  Journal of Power Sources 124~(1) (2003) 299--302.

\bibitem{skyllas2004kinetics}
M.~Skyllas-Kazacos, Y.~Limantari, Kinetics of the chemical dissolution of
  vanadium pentoxide in acidic bromide solutions, Journal of applied
  electrochemistry 34~(7) (2004) 681--685.

\bibitem{wang1984study}
Y.~Wang, M.~Lin, C.~Wan, A study of the discharge performance of the ti/fe
  redox flow system, Journal of power sources 13~(1) (1984) 65--74.

\bibitem{wang2013recent}
W.~Wang, Q.~Luo, B.~Li, X.~Wei, L.~Li, Z.~Yang, Recent progress in redox flow
  battery research and development, Advanced Functional Materials 23~(8) (2013)
  970--986.

\bibitem{rychcik1988characteristics}
M.~Rychcik, M.~Skyllas-Kazacos, Characteristics of a new all-vanadium redox
  flow battery, Journal of power sources 22~(1) (1988) 59--67.

\bibitem{luo2008preparation}
Q.~Luo, H.~Zhang, J.~Chen, D.~You, C.~Sun, Y.~Zhang, Preparation and
  characterization of nafion/speek layered composite membrane and its
  application in vanadium redox flow battery, Journal of Membrane Science
  325~(2) (2008) 553--558.

\bibitem{jiang2020high}
H.~Jiang, J.~Sun, L.~Wei, M.~Wu, W.~Shyy, T.~Zhao, A high power density and
  long cycle life vanadium redox flow battery, Energy Storage Materials 24
  (2020) 529--540.

\bibitem{skyllas2019performance}
M.~Skyllas-Kazacos, Performance improvements and cost considerations of the
  vanadium redox flow battery, ECS Transactions 89~(1) (2019) 29.

\bibitem{Chen2021}
Y.~Chen, Z.~Xu, C.~Wang, J.~Bao, B.~Koeppel, L.~Yan, P.~Gao, W.~Wang,
  {Analytical modeling for redox flow battery design} (2021).
\newblock \href {https://doi.org/10.1016/j.jpowsour.2020.228817}
  {\path{doi:10.1016/j.jpowsour.2020.228817}}.

\bibitem{You2009a}
D.~You, H.~Zhang, J.~Chen, {A simple model for the vanadium redox battery},
  Electrochimica Acta 54~(27) (2009) 6827--6836.
\newblock \href {https://doi.org/10.1016/j.electacta.2009.06.086}
  {\path{doi:10.1016/j.electacta.2009.06.086}}.

\bibitem{Shah2011}
A.~A. Shah, R.~Tangirala, R.~Singh, R.~G. Wills, F.~C. Walsh, {A dynamic unit
  cell model for the all-vanadium flow battery}, Journal of the Electrochemical
  Society 158~(6) (2011) 10--13.
\newblock \href {https://doi.org/10.1149/1.3561426}
  {\path{doi:10.1149/1.3561426}}.

\bibitem{Chen2014a}
C.~L. Chen, H.~K. Yeoh, M.~H. Chakrabarti,
  \href{http://dx.doi.org/10.1016/j.electacta.2013.12.074}{{An enhancement to
  Vynnycky's model for the all-vanadium redox flow battery}}, Electrochimica
  Acta 120 (2014) 167--179.
\newblock \href {https://doi.org/10.1016/j.electacta.2013.12.074}
  {\path{doi:10.1016/j.electacta.2013.12.074}}.
\newline\urlprefix\url{http://dx.doi.org/10.1016/j.electacta.2013.12.074}

\bibitem{Eapen2019}
D.~E. Eapen, S.~R. Choudhury, R.~Rengaswamy,
  \href{https://doi.org/10.1016/j.apsusc.2018.02.025}{{Low grade heat recovery
  for power generation through electrochemical route: Vanadium Redox Flow
  Battery, a case study}}, Applied Surface Science 474 (2019) 262--268.
\newblock \href {https://doi.org/10.1016/j.apsusc.2018.02.025}
  {\path{doi:10.1016/j.apsusc.2018.02.025}}.
\newline\urlprefix\url{https://doi.org/10.1016/j.apsusc.2018.02.025}

\bibitem{zhou2019nano}
X.~Zhou, X.~Zhang, Y.~Lv, L.~Lin, Q.~Wu, Nano-catalytic layer engraved carbon
  felt via copper oxide etching for vanadium redox flow batteries, Carbon 153
  (2019) 674--681.

\bibitem{mayrhuber2014laser}
I.~Mayrhuber, C.~Dennison, V.~Kalra, E.~Kumbur, Laser-perforated carbon paper
  electrodes for improved mass-transport in high power density vanadium redox
  flow batteries, Journal of Power Sources 260 (2014) 251--258.

\bibitem{li2021symmetry}
X.~Li, P.~Gao, Y.-Y. Lai, J.~D. Bazak, A.~Hollas, H.-Y. Lin, V.~Murugesan,
  S.~Zhang, C.-F. Cheng, W.-Y. Tung, et~al., Symmetry-breaking design of an
  organic iron complex catholyte for a long cyclability aqueous organic redox
  flow battery, Nature Energy 6~(9) (2021) 873--881.

\bibitem{LIU2005270}
J.-W. Liu, L.-F. Jiao, H.-T. Yuan, Y.-J. Wang, Q.~Liu,
  \href{https://www.sciencedirect.com/science/article/pii/S0925838805002823}{Effect
  of discharge cut off voltage on cycle life of mgni-based electrode for
  rechargeable ni-mh batteries}, Journal of Alloys and Compounds 403~(1) (2005)
  270--274.
\newblock \href {https://doi.org/https://doi.org/10.1016/j.jallcom.2005.03.069}
  {\path{doi:https://doi.org/10.1016/j.jallcom.2005.03.069}}.
\newline\urlprefix\url{https://www.sciencedirect.com/science/article/pii/S0925838805002823}

\bibitem{cheng2020data}
Z.~Cheng, K.~M. Tenny, A.~Pizzolato, A.~Forner-Cuenca, V.~Verda, Y.-M. Chiang,
  F.~R. Brushett, R.~Behrou, Data-driven electrode parameter identification for
  vanadium redox flow batteries through experimental and numerical methods,
  Applied Energy 279 (2020) 115530.

\bibitem{Sharma2015}
A.~K. Sharma, C.~Y. Ling, E.~Birgersson, M.~Vynnycky, M.~Han, {Verified
  reduction of dimensionality for an all-vanadium redox flow battery model},
  Journal of Power Sources 279 (2015) 345--350.
\newblock \href {https://doi.org/10.1016/j.jpowsour.2015.01.019}
  {\path{doi:10.1016/j.jpowsour.2015.01.019}}.

\bibitem{Lee2019c}
S.~B. Lee, K.~Mitra, H.~D. Pratt, T.~M. Anderson, V.~Ramadesigan, B.~R.
  Chalamala, V.~R. Subramanian, {Open data, models, and codes for vanadium
  redox batch cell systems: A systems approach using zero-dimensional models},
  Journal of Electrochemical Energy Conversion and Storage 17~(1) (2019).
\newblock \href {https://doi.org/10.1115/1.4044156}
  {\path{doi:10.1115/1.4044156}}.

\bibitem{Knehr2011}
K.~W. Knehr, E.~C. Kumbur,
  \href{http://dx.doi.org/10.1016/j.elecom.2011.01.020}{{Open circuit voltage
  of vanadium redox flow batteries: Discrepancy between models and
  experiments}}, Electrochemistry Communications 13~(4) (2011) 342--345.
\newblock \href {https://doi.org/10.1016/j.elecom.2011.01.020}
  {\path{doi:10.1016/j.elecom.2011.01.020}}.
\newline\urlprefix\url{http://dx.doi.org/10.1016/j.elecom.2011.01.020}

\bibitem{newman2012electrochemical}
J.~Newman, K.~E. Thomas-Alyea, {Electrochemical systems}, John Wiley {\&} Sons,
  2012.

\bibitem{Bird2002}
R.~B. Bird, {Transport phenomena} (2002).
\newblock \href {https://doi.org/10.1115/1.1424298}
  {\path{doi:10.1115/1.1424298}}.

\bibitem{Bao2019}
J.~Bao, V.~Murugesan, C.~J. Kamp, Y.~Shao, L.~Yan, W.~Wang, {Machine Learning
  Coupled Multi-Scale Modeling for Redox Flow Batteries}, Advanced Theory and
  Simulations 3~(2) (2020) 1--13.
\newblock \href {https://doi.org/10.1002/adts.201900167}
  {\path{doi:10.1002/adts.201900167}}.

\bibitem{Byrd1995}
R.~H. Byrd, P.~Lu, J.~Nocedal, C.~Zhu, {A Limited Memory Algorithm for Bound
  Constrained Optimization}, SIAM Journal on Scientific Computing (1995).
\newblock \href {https://doi.org/10.1137/0916069} {\path{doi:10.1137/0916069}}.

\bibitem{Kingma2015}
D.~P. Kingma, J.~{Lei Ba}, \href{https://arxiv.org/pdf/1412.6980.pdf %22 entire
  document}{{Adam: A Method for Stochastic Optimization}}, Iclr (2015)
  1--15\href {http://arxiv.org/abs/1412.6980v9} {\path{arXiv:1412.6980v9}}.
\newline\urlprefix\url{https://arxiv.org/pdf/1412.6980.pdf %22 entire document}

\bibitem{Glorot2010}
X.~Glorot, Y.~Bengio, {Understanding the difficulty of training deep
  feedforward neural networks}, in: 13th International Conference on Artificial
  Intelligence and Statistics, 2010.
\newblock \href {http://arxiv.org/abs/arXiv:1011.1669v3}
  {\path{arXiv:arXiv:1011.1669v3}}, \href {https://doi.org/10.1.1.207.2059}
  {\path{doi:10.1.1.207.2059}}.

\bibitem{yang2021delving}
Y.~Yang, K.~Zha, Y.-C. Chen, H.~Wang, D.~Katabi, Delving into deep imbalanced
  regression, arXiv preprint arXiv:2102.09554 (2021).

\end{thebibliography}

\end{document}